# Development of a Highly Selective First-Level Muon Trigger for ATLAS at HL-LHC Exploiting Precision Muon Drift-Tube Data

S. Abovyan, V. Danielyan, M. Fras, P. Gadow, O. Kortner*, S. Kortner, H. Kroha*, F. Müller, S. Nowak, R. Richter, K. Schmidt-Sommerfeld

Max-Planck-Institut für Physik, Föhringer Ring 6, D-80805 München, Germany

**Abstract**

The High-Luminosity LHC (HL-LHC) will provide the unique opportunity to explore the nature of physics beyond the Standard Model of strong and electroweak interactions. Highly selective first-level triggers are essential for the physics programme of the ATLAS experiment at HL-LHC, where the instantaneous luminosity will exceed the instantaneous LHC Run 1 luminosity by about an order of magnitude. The ATLAS first-level muon trigger rate is dominated by low momentum muons, which are accepted because of the moderate momentum resolution of the RPC and TGC trigger chambers. This limitation can be overcome by exploiting the data of the precision Muon Drift-Tube (MDT) chambers in the first-level trigger decision. This requires continuous fast transfer of the MDT hits to the off-detector trigger logic and fast track reconstruction algorithms. The reduction of the muon trigger rate achievable with the proposed new trigger concept, the performance of a novel fast track reconstruction algorithm, and the first hardware demonstration of the scheme with muon testbeam data taken at the CERN Gamma Irradiation Facility are discussed.

## I. CONCEPT OF AN MDT-BASED FIRST-LEVEL MUON TRIGGER

The ATLAS experiment at the Large Hadron Collider [1] currently uses a three-level trigger system. The first-level high transverse momentum ($p_T$) muon trigger is based on trigger chambers with excellent time resolution of better than 20 ns which are able to associate muon tracks to a particular proton bunch crossing. The trigger chambers also provide fast muon momentum measurement, but with limited accuracy due to their moderate spatial resolution. The limited momentum resolution of the trigger chambers weakens the selectivity of the first-level single-muon trigger with a typical threshold of $p_T > 20$ GeV at the High-Luminosity LHC (HL-LHC) by accepting muons below the threshold. Limiting the muon trigger rates rates at HL_LHC requires significantly improved muon momentum resolution of the first-level muon trigger system sharpening the trigger turn-on curve of the trigger as a function of $p_T$. This can be achieved without the installation of new trigger chambers with higher spatial resolution by exploiting the much higher spatial resolution of the Muon Drift-Tube (MDT) tracking chambers of the ATLAS muon spectrometer [2].

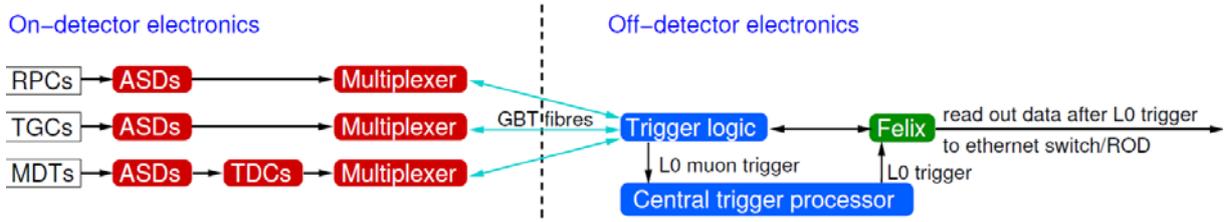

Figure 1: Schematics of the muon chamber data stream in the ATLAS experiment at HL-LHC (see [3]). Trigger chambers (RPC and TGC) and precision muon tracking chambers (MDT) send all their data in a continuous stream to the trigger logic outside the ATLAS detector where they are buffered. The data are read out after a first-level (L0) trigger by so called Felix modules.

In this concept adopted by the ATLAS collaboration (illustrated in Figure 1) [3], trigger (RPC and TGC) and precision tracking chambers (MDT) are continuously read out in trigger-less mode via optical Gbit links and buffered in the first-level muon trigger logic unit until a first-level trigger signal arrives. The trigger chambers provide regions of interest (RoI) containing high-$p_T$ muon candidates. MDT hits inside the RoIs are processed by the muon trigger logic performing track reconstruction and accurate muon momentum measurement which leads to a vastly sharpened trigger $p_T$ turn-on curve and selectivity (see Figure 2).

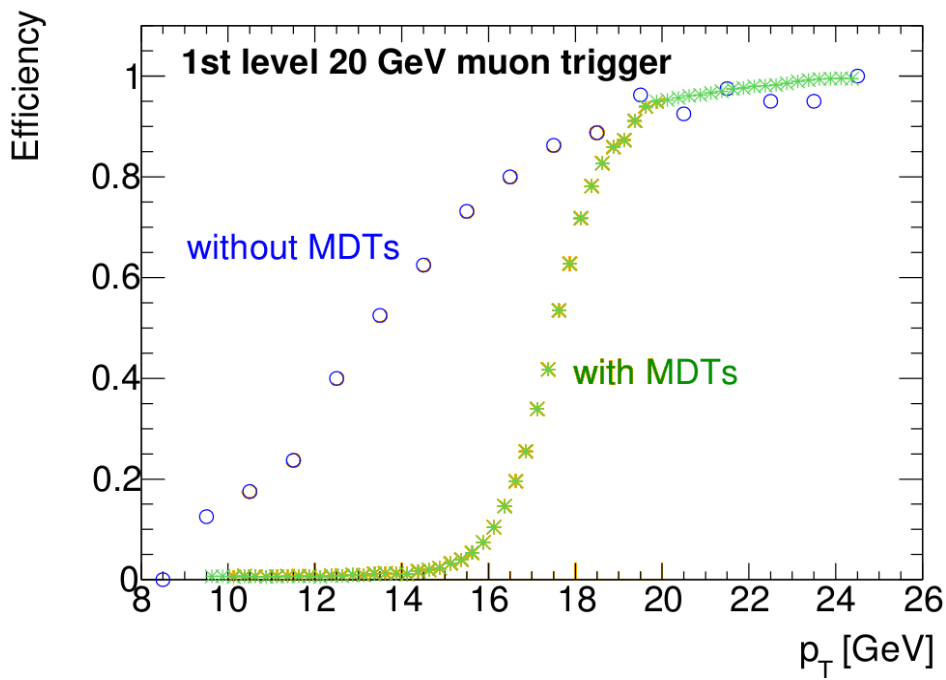

Figure 2: Efficiency of the first-level muon trigger with and without exploitation of the MDT chamber data [4].

## II. PERFORMANCE STUDIES

Studies with Run 1 collision data and with simulated events [4] proved that the proposed MDT based trigger scheme can cope with these high background rates. Figure 2 shows the turn-on of the first-level muon trigger with nominal $p_T$ threshold of 20 GeV with and without the use of the MDT chambers. Exploitation of the MDT data vastly improves the sharpness of the trigger turn-on curve and the selectivity leading to an overall trigger rate reduction by a factor of 5 to about 20 kHz (see Figure 3).

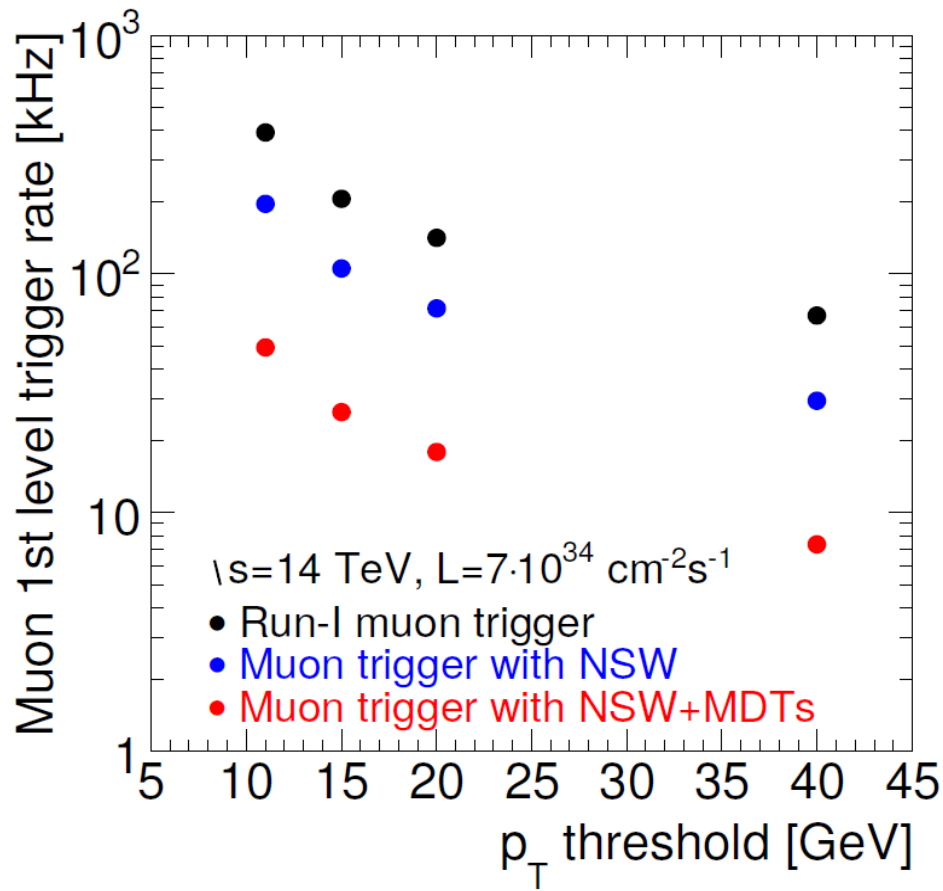

Figure 3: Expected first-level single-muon trigger rates at the HL-LHC design luminosity as a function of the $p_T$ threshold with and without exploitation of the MDT chamber data. New endcap inner layers with increased spatial resolution of the trigger chambers (New Small Wheels, NSW [5]) are foreseen to reduce the endcap muon trigger rates already after the long LHC shutdown 2019/20.

## III. HARDWARE DEMONSTRATOR

A major difficulty in the ATLAS muon system is the presence of large background of thermal neutrons and γ rays causing occupancies of up to 10% in the MDT chambers at HL-LHC. A demonstrator for the trigger-less streamed MDT readout using Xilinx Vertex-6 FPGAs for sending and receiving the hit data was successfully tested in a muon beam at the CERN Gamma Irradiation Facility (GIF++2). No data were lost by the priority readout chain (see Figures 5 and 6). The data were passed to the embedded 1 GHz Cortex-A9 ARM CPU on the receiving end for track reconstruction. A novel fast track reconstruction algorithm based on Hough transformation was implemented in assembler language on the Neom ARM SIMD engine and applied to the GIF data. The distributions of the processing times of the muon tracks for different γ background levels at the GIF are shown in Figure 7. The processing time is less than 3.5 μs which is within the latency of the ATLAS first-level trigger planned for HL-LHC [2]. Further reduction of the processing time and improvements of the MDT trigger efficiency and fake trigger rate are under study (see [4]).

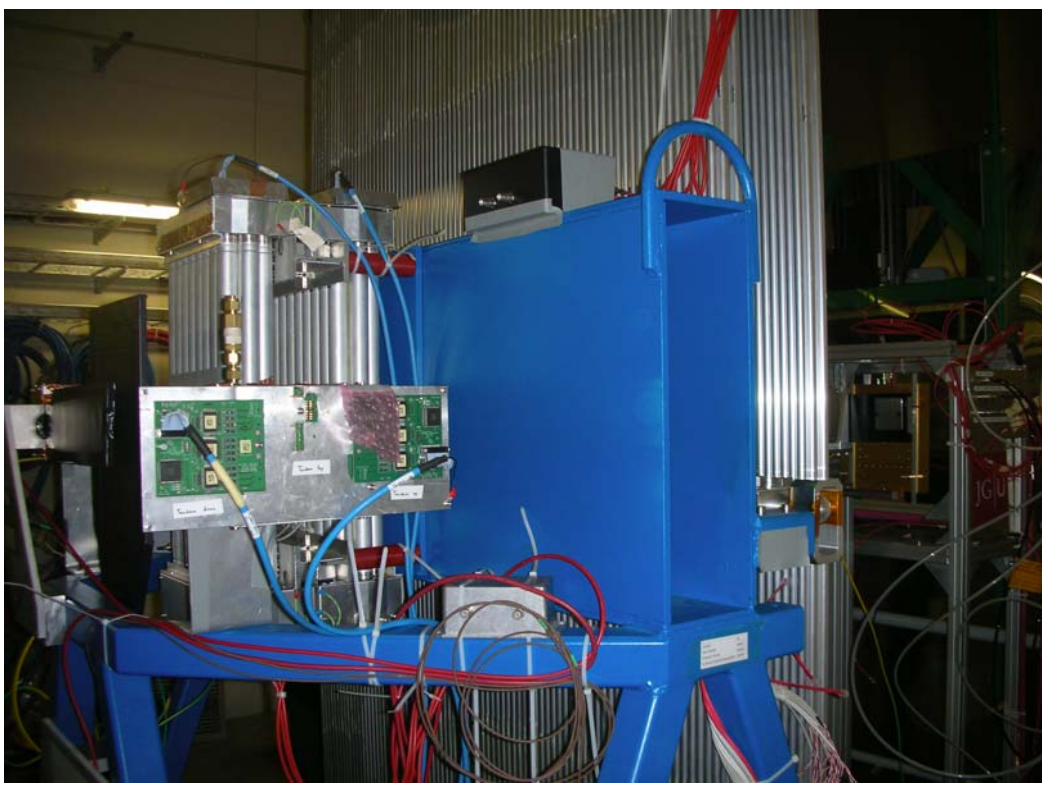

Figure 4: MDT trigger demonstrator setup in the muon beam at the new CERN Gamma Irradiation Facility GIF++ in 2015. The test MDT chamber with two triple layers of 30 cm long drift tubes of 30 mm with the two readout electronics boards containing the FPGA-based continuous-readout TDC can be seen on the left. A small-diameter MDT chamber (sMDT) with 15 mm tube diameter and 2.15 m tube length was used as tracking reference in the muon beam.

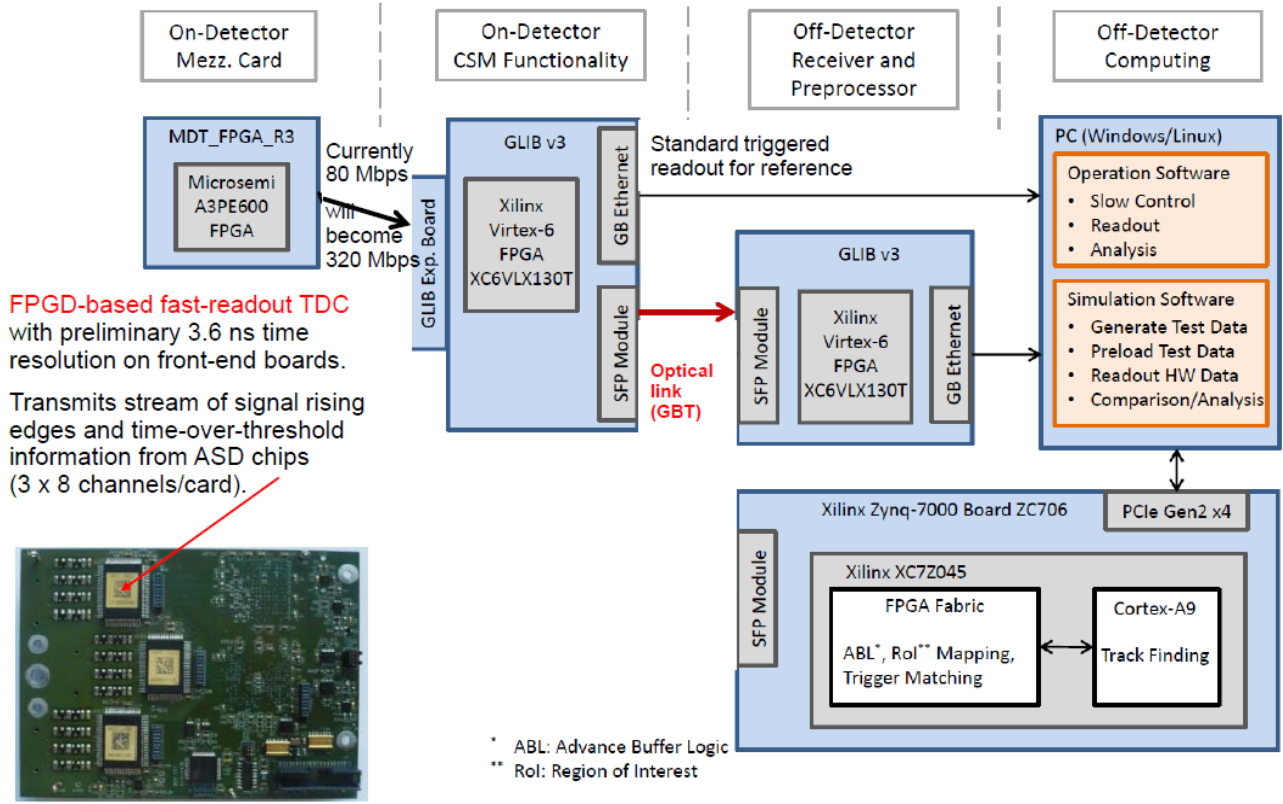

Figure 5: Readout and trigger electronics scheme in the MDT-based muon trigger demonstrator test at the GIF++ in 2015 (see text). The continuous-readout TDC is implemented on the dedicated front-end boards together with the standard MDT TDC for comparison of the drift time measurements. Hit matching with the external scintillator trigger in the muon beam was performed on the receiving Xilinx Virtex-6 FPGA.

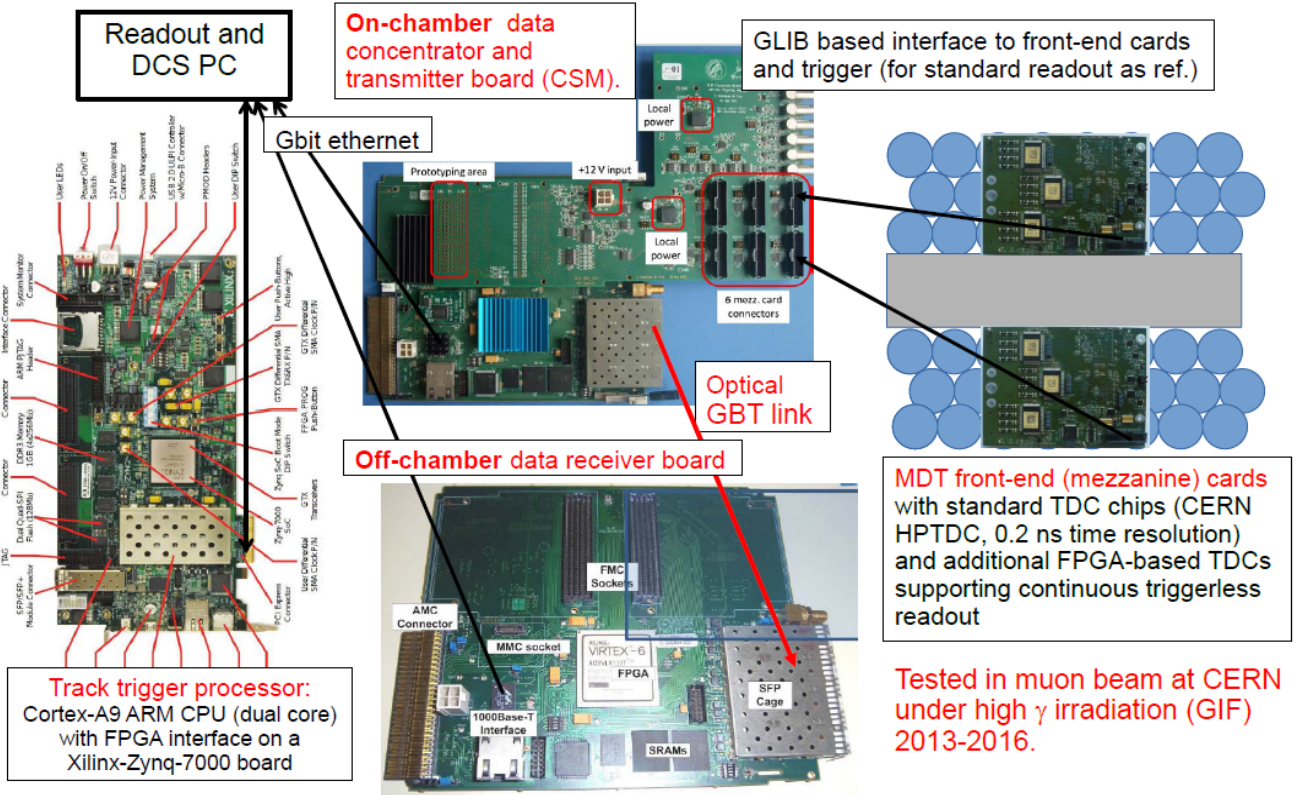

Figure 6: Hardware implementation of the MDT-based first-level muon trigger demonstrator scheme shown in Figure 5.

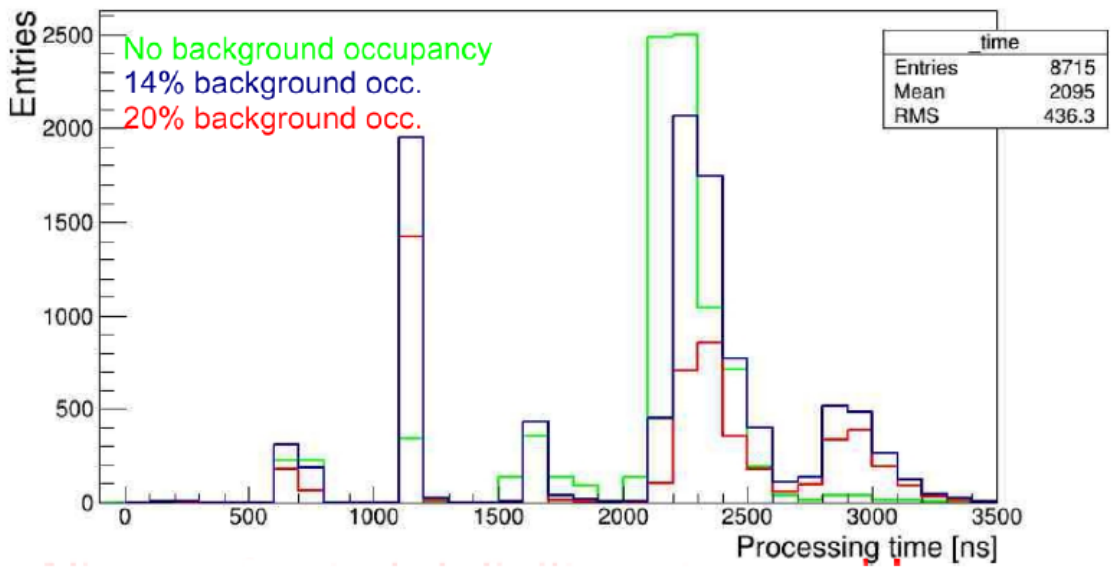

Figure 7: Processing time of a Hough-transform based MDT muon track reconstruction algorithm implemented on a 1 GHz Cortex-A9 ARM CPU for data taken with a MDT chamber at the CERN Gamma Irradiation Facility at different background irradiation and corresponding drift-tube occupancy levels.

# REFERENCES

[1] ATLAS collaboration, G. Aad et al., The ATLAS Experiment at the CERN Large Hadron Collider, JINST 3 (2008) S08003.
[2] J. Dubbert, O. Kortner, S. Kortner, H. Kroha, R. Richter, Upgrade of the ATLAS Muon Trigger for the SLHC, JINST 5 (2010) C12016.
[3] ATLAS collaboration, G. Aad et al., Letter of Intent for the Phase-II Upgrade of the ATLAS Experiment, CERN-LHCC-2012-022, December 2012; ATLAS Phase-II Upgrade Scoping Document, CERN-LHCC-2015-020, September 2015.
[4] Ph. Gadow, Development of a Concept for the Muon Trigger of the ATLAS Detector at the HL-LHC, PhD thesis, Max-Planck-Institut für Physik, Munich, and Technical University Munich, MPP-2016-086, CERN-THESIS-2016-056, June 2016.
[5] ATLAS collaboration, G. Aad et al., New Small Wheel Technical Design Report, CERN-LHCC-2013-006, June 2013.